\documentclass[aps,prl,twocolumn,groupedaddress]{revtex4}
\usepackage[dvips]{epsfig}

\begin{document}

\title{The ordering temperature and critical exponents of the binomial Ising spin glass in dimension three}
\author{P. O. Mari}
\address{Laboratoire de Physique des Solides,
\\ Universit\'e Paris Sud, 91405 Orsay, France\\}
\author{I. A. Campbell}
\address{Laboratoire des Verres, \\ Universit\'e Montpellier II,
       34095 Montpellier, France}
\date{\today}

\begin{abstract}

We compare numerical estimates from different sources for the ordering temperature $T_g$ and the critical exponents of the Ising spin glass in dimension three with binomial ($\pm J$) interactions. Corrections to finite size scaling turn out to be important especially for parameters such as the Binder cumulant. For non-equilibrium parameters it is easier to approach the large size limit and to allow for corrections to scaling. Relying  principally on such data, a crossing point defines the freezing temperature $T_g$; the possibility that the ordering temperature is zero can definitively be excluded. We estimate an ordering temperature $T_g = 1.195(15)$, with associated estimates of the critical exponents for which corrections to finite size scaling are well under control. Among the parameters evaluated is the leading dynamic correction to scaling exponent $w$.

\end{abstract}

\maketitle
\section {Introduction}

The Edwards-Anderson Ising spin glass (ISG) has been studied intensively for over twenty five years. In this model Ising spins on a regular lattice in dimension three are coupled by random near neighbour interactions \cite{EA}. Despite its deceptively simple definition, the basic physics of this canonical complex system still eludes consensus. Recently interest has focused on the correct description of the ground state and the low energy excited states \cite{ground}, but finite temperature aspects are also highly important, in particular the critical behaviour at the freezing temperature $T_g$. Over the years a number of numerical efforts have been consecrated to accurate measurements near $T_g$, with the aim first of demonstrating that $T_g$ is non-zero, and then of determining the critical exponents precisely \cite{BY,og,KY,pala,pom,Bal}. The long relaxation times and the need to average over large numbers of independent samples renders the task numerically laborious. In addition, it has not always been appreciated that corrections to finite size scaling can be singularly vicious in this system.

 Here we will concentrate on the problem of the accurate and reliable determination of $T_g$ and the critical exponents for the ISG with binomial ($\pm J$) near neighbour interactions in a dimension three simple cubic lattice ($L\times L\times L$),

\begin{equation}
H= - \sum J_{i,j}S_iS_j
\end{equation}
with $J_{i,j}$ chosen at random as $\pm 1$. As this system is the dropsophilia of spin glass studies, it is important to establish as precisely and reliably as possible the basic parameters of its critical behaviour. We will show that a parameter set can be obtained  consistent with most published data from different complementary approaches, if corrections to finite size scaling are properly taken into account.

We will principally refer to the simulations of references \cite{og,KY,pala,pom,Bal}; the estimates drawn from the analyses of the numerical data are shown in Table I. We will comment on these data as we go along. The main numerical contribution of the present work is to present non-equilibrium data of higher statistical accuracy than before, and to allow for corrections to finite size scaling in the analysis of these results also.

As a general rule, the apparent exponent values are strongly correlated with the value assumed for the ordering temperature, so once $T_g$ is well determined it is relatively easy to obtain good values for the exponents.  On the contrary, if the wrong $T_g$ is used, the estimates of the exponents will be biased. Finite size scaling methods have been widely used, but corrections to finite size scaling have often been overlooked, which is dangerous. In Table I the estimates of the ordering temperatures and of the exponents from different publications should not be considered as independent of each other, but each parameter set should be taken as a whole.

The autocorrelation function at time $t$ is defined by
\begin{equation}
q(t) = <S_i(0).S_i(t)>
\end{equation}
where the average ($<..>$)is over all spins. In practice a large number of further averages (denoted by $[..]$)are taken over many samples all of size $L^3$ spins with microscopically different arrangements of the bonds. We will use $q$ to mean $q(t)$ at times long enough for the system to have entered the equilibrium fluctuation regime. Periodic (toroidal) boundary conditions are generally but not always used (see \cite{Bal}).
The equilibrium spin glass susceptibility is given by
\begin{equation}
\chi_L(T)= L^3 [<q^2>]
\end{equation}
where the average is taken over time once the sample has been annealed to equilibrium either by carrying out enough annealing steps or by using the more sophisticated and efficient parallel tempering method \cite{KY,Bal}.
The non-equilibrium susceptibility $\chi(t)$ is the spin glass susceptibility at time $t$
\begin{equation}
\chi(t)=L^3 [<q^2(t)>]
\end{equation}
after a simple anneal which starts from a totally uncorrelated (infinite temperature) state and lasts a time $t$ \cite{huse}.

\section{Binder parameter}
The first large scale simulations on this problem were those of Ogielski \cite{og} who did massive dynamic and static measurements on samples as large as $L=64$. His values were considered authoritative until finite size scaling techniques were later intensively exploited \cite{BY,KY,pala,Bal}.
The standard scaling method to estimate $T_g$ from finite size data in ISGs is through the use of the dimensionless Binder parameter,

\begin{equation}
g_L(T) = [3 - <q^4>/<q^2>^2]/2
\end{equation}

where $<q^2>$ and $<q^4>$ are moments of the fluctuations of the autocorrelation parameter $q(t)$ in thermal equilibrium at temperature $T$. The averages are taken first over time and then over samples. In absence of corrections to finite size scaling, the family of curves $g_{L}(T)$ for different sizes $L$ must all intersect precisely at $g_{0}$ for $T=T_g$. For the 3d binomial ISG, the $g_L(T)$ curves come together near $T=1.20$ \cite{BY} but very good statistics are needed to see intersections \cite{KY,Bal}. Precisely because the $g_L(T)$ curves lie very close together, even small corrections to finite size scaling can have a drastic effect on the crossing point temperatures. As corrections to finite size scaling are already visible for the spin glass susceptibility \cite{pala,pom}, corrections can be expected {\it a fortiori} for the Binder parameter. If $\omega$ is the leading correction to scaling exponent, at $T_g$ the successive $g_L$ values for increasing $L$ can be expected to behave as
\begin{equation}
g_L = g_0[1 + const* L^{-\omega}]
\end{equation}

It has been pointed out \cite{Bal} that in these circumstances it is helpful to use the "quotient method". An elementary version of this method is to use "crossing points".
Successive intersection temperatures $T^{*}(L)$, corresponding to intersections of curves $g_L(T)$ and  $g_{2L}(T)$, are plotted. If the leading correction exponent $\omega$ dominates, then $T^{*}(L)$ values should extrapolate to the true infinite $L$ ordering temperature $T_g$ as

\begin{equation}
T^{*}(L) = T_g + const* L^{-(\omega+1/\nu)}
\end{equation}

Extrapolation to infinite $L$ provides a precise estimate for $T_g$, if the corrections to finite size scaling are fully under control. In the case of the 3d binomial ISG, accurate $g_L(T)$ data for sizes $L=5,10, 20$ \cite{Bal}, for $L= 6,8,12,16, 24$ \cite{KY} and for $L=3,4$ \cite{pom} can be used to obtain intersections. It should be noted that for technical reasons the boundary conditions used by Ballesteros et al \cite{Bal} on a specially built dedicated computer are helicoidal while for the other calculations the boundary conditions are periodic. A direct comparison of the raw data from \cite{cruz} and \cite{KY} indicates that systematic differences in the results because of the different boundary conditions are negligible for the spin glass susceptibility at $L=20$. The sequence of Binder parameter curves for $L=3,4,5,6,8$ \cite{pom,Bal,KY} behaves smoothly, although the $L=5$ curve is the only one in the set to have helicoidal boundary conditions. A subset of the data from the different sources is shown for illustration in figure 1. 

\begin{figure}
\begin{center}
\epsfig{file=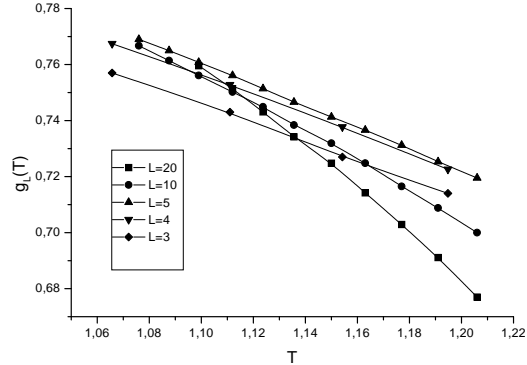,angle=0,width=8.4cm}
\end{center}
\caption{A selection of results for the Binder parameter $g_L(T)$ as function of the temperature $T$. Data for $L=5,10,20$ from \protect\cite{Bal}, data for $L=3,4$ present work. Error bars are about the size of the symbols.}
\label{figure:1}
\end{figure}

\begin{figure}
\begin{center}
\epsfig{file=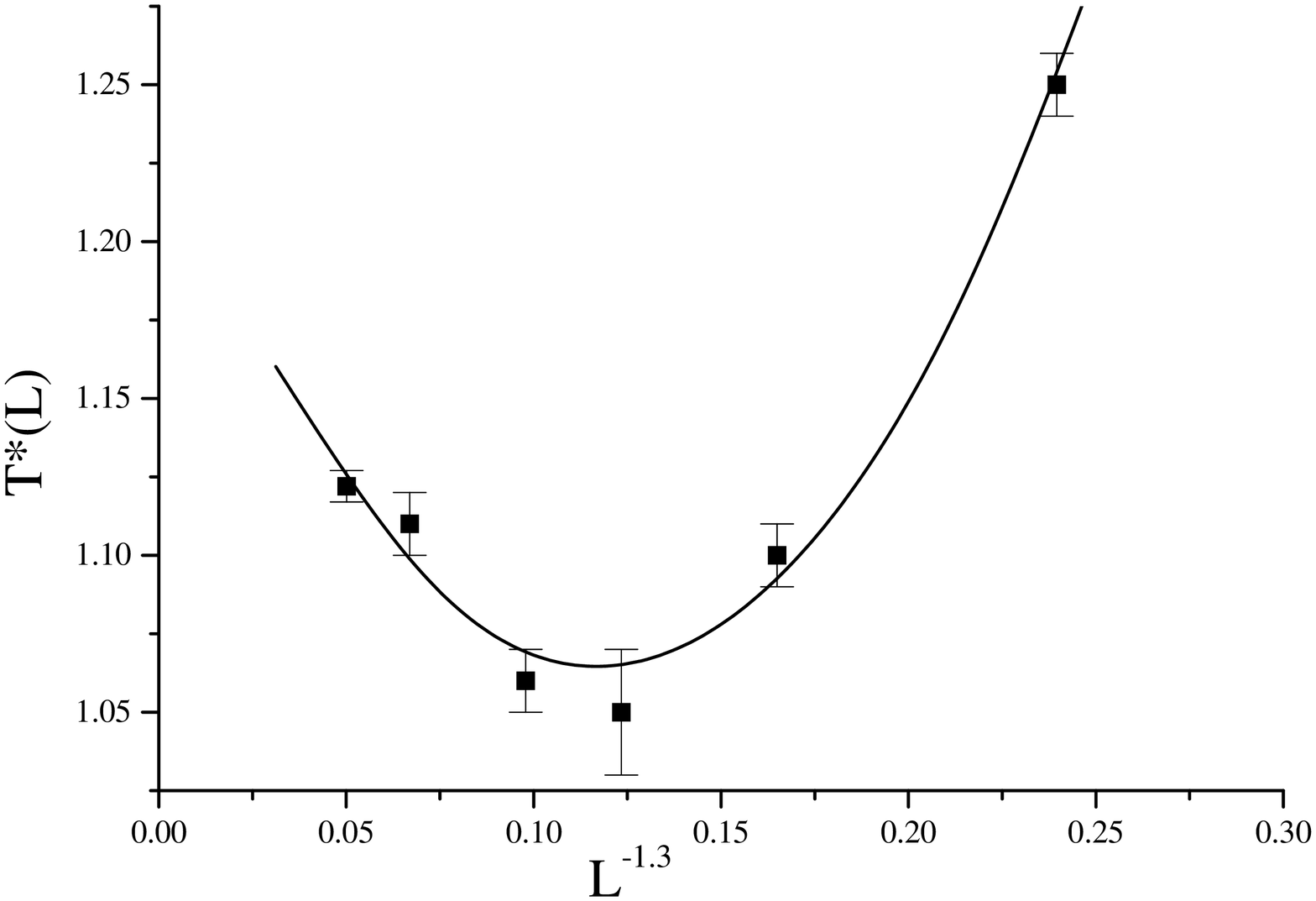,angle=0,width=8.4cm}
\end{center}
\caption{Quotient plot of crossing points for the Binder parameter. $T_{cross}(L)$ is the temperature at which the curves $g_L(T)$ and $g_{2L}(T)$ cross. The $x$ axis is $L^{-1.3}$ (see text).The curve is just a guide to the eye. $T_{cross}(L)$ should extrapolate linearly to $T_g$ for infinite size $L$ as $L^{-(\omega +1/\nu}$.  }
\label{figure:2}
\end{figure}

 It can be seen that the behaviour of the $g_L(T)$ curves and hence of the crossing temperatures varies in a non-monotonic manner with $L$. (We suggest below that this may be because $\omega$ is high and analytic corrections may be non-negligible). The $\omega$ and $\nu$ values of  Ballesteros et al \cite{Bal} correspond to an effective exponent $(\omega+1/\nu)=1.3(4) $. (Estimates of $\omega$ vary considerably from author to author and we will return to this point later).  In figure 2 we have plotted $T^{*}(L)$ in the form of equation (2) using the central value of the exponent. The pairs of sizes $[2L,L]$ we have used are [20,10] and [10,5] where the boundary conditions are helicoidal, [16,8],[12,6],[8,4] and [6,3] where the boundary conditions are periodic. There is no very obvious systematic difference between the two sets of points with different boundary conditions. Even for the larger sizes the crossing temperature is still increasing fairly rapidly with increasing $L$ ; the very high quality data of \cite{Bal} show a ${L=10,20}$ crossing already at $T^{*}=1.118(10)$, so the central estimate given by \cite{KY}, $T_g =1.11$, is definitely too low.  The curve in figure 2 (a two paramereter spline fit) is just a schematic indication of one possible extrapolation to to infinite $L$ which would appear to lead to a significantly higher estimate for $T_g$ than that of \cite{Bal}, $T_g = 1.138(10)$. Alternative curves could be drawn through the data points, but the Binder parameter data (including the high quality results of \cite{Bal}) point to a higher $T_g$ than their central estimate. The non-monotonic behaviour of the crossing points means that not only the leading correction term is operative. Thus in the case of this particular ISG system the Binder parameter method can only give a rough estimate of $T_g$, because the correction terms are badly under control.

It should be noted that because the behaviour of the crossing temperature $T^{*}(L)$ is not monotonic, if only measurements of $g_L(T)$ for $L = 4, 8,$ and $16$ were available one would see an excellent unique intersection of the three $g_L(T)$ curves, but this intersection point is NOT at the true infinite size limit temperature $T_g$. It is an artefact of the non-monotonic behaviour of $T^{*}(L)$.

Binder parameter data can also be directly used to estimate the exponent $\nu$, as
\begin{equation}
d(g_L)/d(1/T) \propto L^{1/\nu}
\end{equation}
at $T_g$. Using a single $g_L(T)$ data set, the value of $\nu$ estimated in this way depends strongly on the $T_g$ estimate. The variation of the $g_L(T)$ slopes with $L$ in the $g_L$ data of Ballesteros et al provide an apparent $\nu$ value which is about $2.15$ if $T_g = 1.138$ (their preferred value); if $T_g$ is higher then the associated estimate for $\nu$ from the same data set is lower. Once again estimates can be affected also by correction terms. (Similarly, the apparent $\eta$ from spin glass susceptibility data changes rapidly with the assumed $T_g$, Figure 7).

\section{Correlation length}
Ballesteros et al \cite{Bal} do not in fact base their $T_g$ estimate on their Binder parameter data, but use an independent criterion : the intersection of curves for the ratio of the size dependent correlation length to $L$. However they show data for three sizes only ($L=5,10,20$) and do not discuss possible corrections to finite size scaling for this parameter, even for $L$ as low as $L=5$. Even though their intersection point for these three sizes appears to be unique and very clear, the $T_g$ estimate from this criterion ($T_g = 1.138(10)$) is lower than the prefered $T_g$ value obtained from other measurements which we will discuss below, $T_g = 1.195(15)$. Although both series of measurements show clear evidence for a well established transition, and though the difference between the estimated transition temperatures is only about $4\%$, it would be more satisfactory if complete agreement could be reached.

\section{Series}

It is of interest to note that quite independent information from series calculations can throw further light on the values of the ordering temperature and the exponents. Long series results on the 3d binomial system  \cite{singh} provide sets of  parameters $[T_g, \gamma]_a$ corresponding to different approximants, Figure 3. The parameter pair  $[T_g, \gamma]$ representing the true phase transition can be expected to lie somewhere within the cloud of approximant points. In Figure 3 we represent the approximant points from \cite{singh} together with the equivalent points from the different numerical simulation estimates from Table I (using
$\gamma=\nu(2-\eta)$). It can be seen that while the approximant cloud has a relatively wide spread in temperature and so is not very discriminatory for $T_g$, the cloud has a fairly narrow distribution in $\gamma$. The $\gamma$ values corresponding to the parameter sets of references \cite{Bal}, \cite{KY}and \cite{pala} in Table I lie systematically above the series data, while the $[T_g, \gamma]$ pairs of \cite{og} and \cite{pom} fall neatly in the centre of the approximant cloud. In dimension 4 the agreement between series \cite{singh} and simulations \cite{d4} is excellent, so there seems no {\it a priori} reason to expect the series results to be strongly biased in dimension 3.  

\begin{figure}
\begin{center}
\epsfig{file=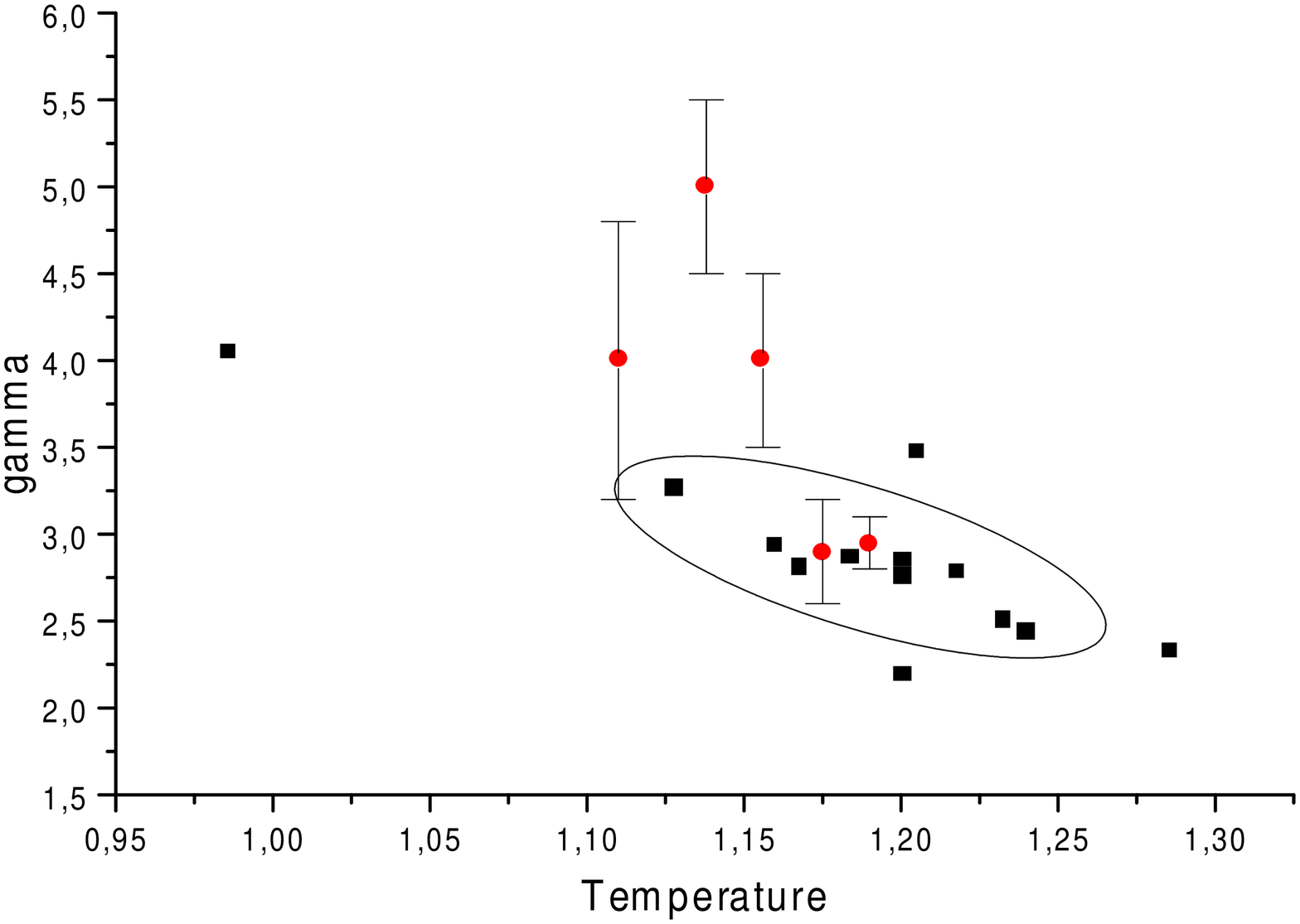,angle=0,width=8.4cm}
\end{center}
\caption{The estimated critical exponent $\gamma$ plotted against the estimated ordering temperature $T_g$. Squares are values obtained for different approximants from series calculations \protect\cite{singh}; the couple $[T_g,\gamma]$ corresponding to the true ordering temperature and exponent for the system can be expected to lie somewhere within the cloud of points indicated by the ellipse. Circles are values reported from different simulations; from left to right : \protect\cite{KY},\protect\cite{Bal},\protect\cite{pala},\protect\cite{og},\protect\cite{pom}. }
\label{figure:3}
\end{figure}

\section{Non-equilibrium methods}

We have proposed a novel technique including non-equilibrium data for identifying $T_g$ numerically through the combination of three simulations \cite{lwb,pom}. We will use it again in the present context in a slightly modified form.

First, the non-equilibrium spin glass susceptibility for samples quenched to $T_g$ increases with time as $t^{h}$ where
\begin{equation}
h=(2-\eta)/z
\end{equation}
$\eta$ and $z$ being the standard equilibrium static and dynamic exponents \cite{huse}.

Secondly the autocorrelation function decay $q(t)$ at $T_g$ is of the form $t^{-x}$ , if a sample is initially annealed at $T_g$ during a long waiting time $t_w$, under the condition that the measuring time for the decay $t$ is much shorter than $t_w$ \cite{og,rieger}. The exponent  $x$ at $T_g$ is related to the equilibrium exponents through $x=(d-2+\eta)/2z$ \cite{og}.

Thirdly, the equilibrium spin glass susceptibility at $T_g$ increases with sample size as \cite{BY}
\begin{equation}
\chi_L/L^2 \propto  L^{-\eta}
\end{equation}

The effective exponent $h(T)$ can be measured directly as a function of trial temperatures $T$. It can also be evaluated indirectly and independently through combining the effective exponent $\eta(T)$ from the finite size scaling of the equilibrium susceptibility, with the effective decay exponent  $x(T)$. We will refer to this derived value as $h^{*}(T)$ :

\begin{equation}
h^{*}(T)=2x(T)(2-\eta(T))/(d-2+\eta(T))
\end{equation}

Measurements of each of the three effective parameters ($h(T),x(T),\eta(T)$) can be made at a series of trial temperatures $T$; consistency dictates that at $T_g$ we must have $h = h^{*}$ so when  $h(T)=h^{*}(T)$, then $T=T_g$.

Once again, care must be taken to avoid errors arising from corrections to finite size scaling. For the measurements of $h(T)$, large samples can readily be used, so it is relatively easy to insure that for the time range used the system is far from the saturation limit for that particular size, see \cite{zheng}. At short times there can however be finite time corrections, the analogue of finite size corrections, as the clusters of correlated spins that are gradually building up with annealing time $t$ have a $t$ dependent finite size, even if the sample size $L$ can treated as effectively infinite. Including the finite time correction, we should write \cite{parisi}
\begin {equation}
\chi(t) = At^{h}[1-Bt^{-w/z}]
\end{equation}
with $h(T)$ as above, and $w$ the leading dynamic correction to scaling exponent (which does not have to be equal to the static exponent $\omega$\cite{ma}). The factors $A$ and $B$ may vary with $T$.
An example of data is shown for two alternative plots in figures 4 and 5. (Rather than $\chi(t)$ we plot $\chi(t)-1$ because at infinite $T$, $\chi(t)=1$. Measurements were averaged over runs on 6766 samples of size $L=28$ at temperature $T=1.2$. For this size saturation to the finite size limited equilibrium susceptibility would only occur at $t  \simeq 10^8$). For this particular temperature the best fit gives for the important exponent $h(T)$ a value of $0.390(5)$. Our results for $h(T)$ at different temperatures, figure 7, are in good agreement with those of \cite{huse,bray}, but have higher statistical accuracy and allow for the correction to scaling which was not considered in the earlier work. We have checked that our estimates for $h(T)$ do not change when we use sample sizes $L$ from $20$ to $28$; this means that we should be in the limit such that the $h(T)$ values represent the infinite size behaviour.

\begin{figure}
\begin{center}
\epsfig{file=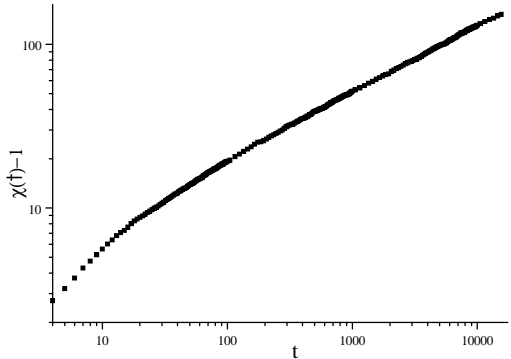,angle=0,width=8.4cm}
\end{center}
\caption{An example of the extraction of the Huse exponent $h(T)$ \protect\cite{huse} from non-equilibrium susceptibility $\chi(t)$ data. $(\chi(t)-1)$ is plotted against time $t$ on a log/log scale. The data in this particular plot correspond to $T=1.2$ and are an average over 6766 independent samples. Data : squares, fit : circles }
\label{figure:4}
\end{figure}

\begin{figure}
\begin{center}
\epsfig{file=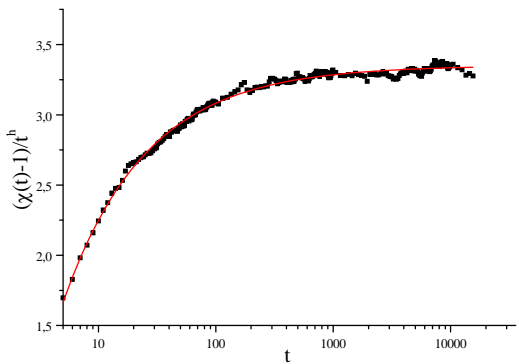,angle=0,width=8.4cm}
\end{center}
\caption{The same data as in Figure 4, but presented in the form of a $(\chi(t)-1)/t^h$ against $log(t)$ plot with $h(T)$ chosen as the best overall fit value. The values of $h, w/z, A$ and $B$ are adjusted to give the best fit to
$(\chi(t)-1)/t^h = A(1 +Bt^{-w/z})$. Data : squares, fit : line }
\label{figure:5}
\end{figure}

Estimates of $w/z$ from the analyses of the non-equilibrium susceptibility data have statistical fluctuations but do not show any systematic trend with $T$ over the fairly narrow range of $T$ values near $T_g$ that we have explored, so our estimate of $w/z$ is not sensitive to the exact choice for $T_g$. The mean value is $w/z = 0.48(10)$. As a number of independent measurements have shown that the dynamic exponent $z$ is near $6$ (\cite{og,bray,pom} and our estimate given below at the end of the analysis), we can deduce from $w/z$ that the dynamic correction to scaling exponent $w = 2.9(6)$. This turns out to be very close to the static exponent value $\omega \sim 2.8$ estimated from an analysis of the size dependence of the spin glass susceptibility $\chi(L)$ \cite{pom}. It is distinctly higher than the estimate $\omega = 0.8(4)$ given by \cite{Bal}. (An attempt to fit the $q^2(t)$ data with a correction exponent $w/z$ of the order of $1/6$ fails completely).

There is an independent check on the expected size of $\omega$. Estimates of $\omega$ can be obtained from series expansion data through $\omega = \Delta_1/\nu$ \cite{klein}
where $\Delta_1$ the appropriate correction to scaling parameter for the series work. In dimension 4 a reliable value $\omega \sim 3.0$ was obtained \cite{klein}; in dimension 3 the estimate by the same authors is $\sim 3.0$ with rather lower accuracy. In the sequence of Ising ferromagnets \cite{pelissetto} and for the sequence of percolation models \cite{perco}, $\omega$ increases regularly with $d_u - d$, where $d_u$ is the upper critical dimension ($4$ for Ising ferromagnets, $6$ for the percolation model as for ISGs). The high value of $\omega$, near $3$ in dimension 3 from the ISG series work, is consistent with the value from the present simulation analysis and is also consistent with there also being a general upward trend for $\omega$ with increasing $d_u - d$ in ISGs. A much lower estimate for $\omega$ \cite{Bal} seems inconsistent with the series work.

A large correction exponent means that the correction factors tend towards $1$ quite rapidly as $L$ or $t$ increase, so the large size or long time limiting behaviour can be reached fairly readily. However for large $\omega$ it may not be safe to ignore analytic corrections to scaling \cite{Bal}, which could perhaps explain the non-monotonic behaviour of $T^{*}_L$ from the Binder parameter data seen above, figure 2.

For the estimates of $x(T)$, the measurement runs for $q(t)$ must be preceded by long annealing runs. It turns out that after an appropriate anneal the algebraic form with constant $x(T)$ sets in from rather short times (a few Monte Carlo steps), and the measurements are self averaging \cite{og,rieger}. Deviations from the asymptotic behaviour are so small that no meaningful analysis in terms of corrections to scaling could be carried out (see for instance the $q(t)$ data shown in \cite{og}). As a consequence it is  relatively easy to obtain robust and accurate values of $x(T)$ on large samples. The data of references \cite{og} and \cite{rieger} are accurate and in excellent agreement with each other, figure 6. Our own data are in full agreement with these values so we have relied on the published measurements. In ISGs the relaxation function continues to be algebraic for temperatures below $T_g$, so there is no problem in defining $x(T)$, while above $T_g$ an additional relaxation term with a cutoff function form appears. Ogielski introduced phenomenologically the stretched exponential decay in this context to parametrise the cutoff function, and it has been frequently used since. For present purposes, at temperatures where the pure algebraic decay no longer holds, one is clearly above $T_g$.

\begin{figure}
\begin{center}
\epsfig{file=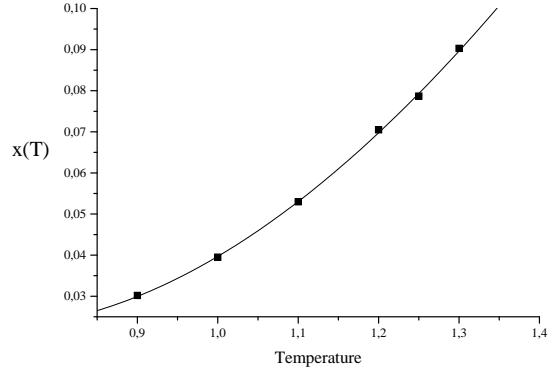,angle=0,width=8.4cm}
\end{center}
\caption{The autocorrelation function initial decay exponent $x(T)$; data from \protect\cite{og,rieger}. }
\label{figure:6}
\end{figure}

Again, as for $h(T)$, $x(T)$ can be measured on almost arbitrarily large samples. This is an important advantage of methods where achieving strict thermal equilibrium is not obligatory.

Numerically the most demanding of the three measurements is that of the size dependent equilibrium spin glass susceptibility $\chi(L,T)$. We have relied on the excellent data of \cite{KY}, which extend up to sizes $L$ from $16$ to $24$ depending on the temperature. The method of allowing for corrections to finite size scaling is outlined in \cite{pom}. Essentially, a reliable estimate of the effective temperature dependent exponent $\eta(T)$ can be obtained from a fit to $\chi(L,T)$ including the correction factor
\begin{equation}
\chi(L,T)/L^2 \propto L^{-\eta(T)}(1+ const L^{-\omega})
\end{equation}
with an effective exponent $\eta(T)$ at and below $T_g$ and a temperature independent $\omega$. The correction term turns out to be more important than the statistical error only for $L < 8$. (We have checked that the raw equilibrium $\chi(L,T)$ data exhibited by \cite{cruz} are in excellent agreement with those of \cite{KY}). The $\chi(L,T)$ results can be taken to provide good infinite size limit estimates for $\eta(T)$, figure 7, with individual errors of about $\pm 0.03$ including statistical and systematic contributions.
Above $T_g$ the $\chi(L,T)/L^2$ curves start to bend down with large $L$, as would be expected from the general scaling form.

\begin{figure}
\begin{center}
\epsfig{file=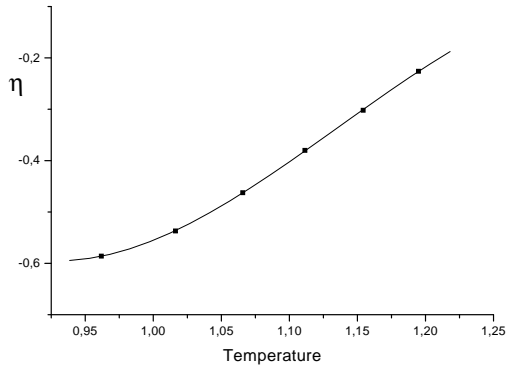,angle=0,width=8.4cm}
\end{center}
\caption{The effective value of the equilibrium exponent $\eta(T)$, eqn (3), from finite size scaling data \protect\cite{KY}. }
\label{figure:7}
\end{figure}

The results for $h(T)$ and $h^{*}(T)$ as defined above are shown in figure 8. There is a clean intersection at $T=1.195(15)$ which we can take to be a reliable estimate for $T_g$, essentially free of corrections to finite size scaling because in each data set care has been taken to use as large samples as possible, and to allow for corrections to finite size scaling whenever appropriate.  The clean intersection should set at rest any lingering suspicion \cite{marinari} that there is not a true transition in this spin glass at a critical temperature $T_g$ close to that estimated by Ogielski \cite{og} many years ago.

\begin{figure}
\begin{center}
\epsfig{file=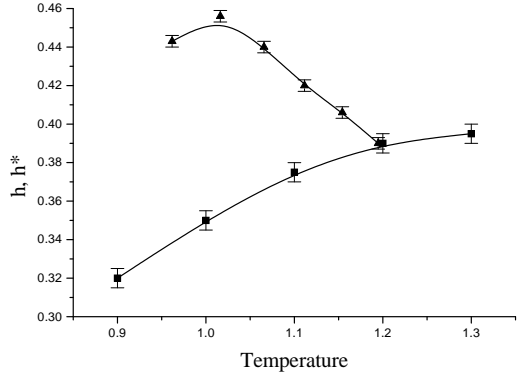,angle=0,width=8.4cm}
\end{center}
\caption{The Huse exponent $h(T)$ measured directly as in Figures 4 and 5 (squares), and the same exponent $h^{*}(T)$ measured via the parameters $x(T),\eta(T)$, eqn (4) (triangles). For consistency, the intersection of the two curves must occur at the ordering temperature $T_g$.}
\label{figure:8}
\end{figure}

We obtain the corresponding exponents from an analysis of the various data, with $T_g$ in hand, Table I. Consistent and precise values are obtained using finite size scaling for $\chi(L,T)$ \cite{KY} with correction terms \cite{pom}, and from the large sample data shown by Ogielski \cite{og}. The value estimated for $z$ directly from the critical values of $h$ and $x$ is $5.65(15)$ which can be compared with $6.0(8)$ from \cite{og}. The major source of error for all the exponents (as reflected in the error bars) is the estimated uncertainty in $T_g$.
\begin{table}
\caption[tableI]{Estimates of $T_g$, $\nu$, $\eta$ and $\gamma$ for the 3d ISG
with binomial interactions using different techniques (see text)}

\label{Table:1}
\begin{center}
\begin{tabular}{ccccc}
Reference                      & $T_g$         & $\nu$         & $\eta$    & $\gamma$ \\
\hline
\\

Ogielski\cite{og}       	& $1.175(25)$   &  $1.3(1)$ & $-0.22(5)$    &  $2.9(3)$   \\
Singh-Chakravarty\cite{singh}   &  $1.18(8)$    &           &               &   $2.8(4)$  \\
Kawashima-Young\cite{KY}        &  $1.11(4)$    &  $1.7(3)$ &   $-0.35(5)$  &   $4.0(8)$  \\
Palassini-Caracciolo\cite{pala} &  $1.156(15)$  & $1.8(2)$  &   $-0.26(4)$  & $4.1(5)$  \\
Mari-Campbell \cite{pom}    &  $1.19(1)$    &   $1.33(5)$   &  $-0.22(2)$   & $2.95(15)$\\
Ballesteros et al\cite{Bal}     &  $1.138(10)$  &  $2.15(15)$  & $-0.337(15)$ & $5.0(4)$\\
Present analysis             &  $1.195(15)$  &  $1.35(10)$  & $-0.225(25)$   & $2.95(30)$ \\

\end{tabular}

\end{center}
\end{table}

Finally, if we plot $\eta(d)$ as a function of dimension $d$ we obtain the curve shown in Figure 9, where we have included the present result (Table 1) for dimension $3$, and values from \cite{jerome}, \cite{endo,ian}, \cite{klein,ian} and the mean field value($\eta = 0$) for dimensions $2,4,5$ and $6$ respectively. (We can note that $\eta$ remains well defined even in dimension 2 where $T_g$ tends to zero \cite{jerome}). The curve is a guide for the eye. The overall shape of $\eta(d)$ is strikingly similar to that for the analagous series of $\eta(d)$ values in the percolation model \cite{adler}.  The present estimate for $\eta(d=3)$ intrapolates smoothly into the sequence of values obtained independently for the other dimensions.

\begin{figure}
\begin{center}
\epsfig{file=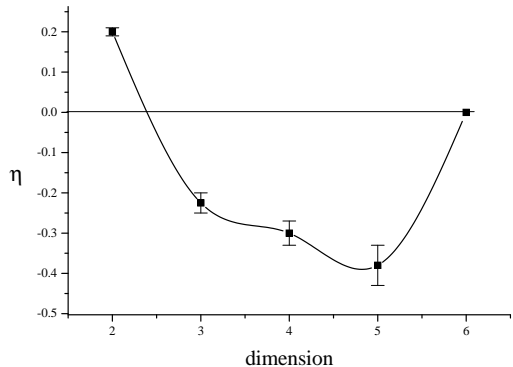,angle=0,width=8.4cm}
\end{center}
\caption{The exponent $\eta(d)$ as a function of dimension $d$ with values taken from references as indicated in the text. The curve is a guide for the eye.}
\label{figure:9}
\end{figure}

\section{Conclusion}

In conclusion, we have once again considered the problem of obtaining reliable estimates for the ordering temperature of the binomial interaction ISG in dimension three and for the associated critical exponents. It is very important to allow for corrections to finite size scaling for this particular system. It is clear from figures 1 and 2 that Binder parameter crossing points in this ISG are strongly affected by substantial corrections to finite size scaling even up to and beyond sizes as large as $L=20$. For the spin glass susceptibility $\chi$, corrections to scaling are also present but their influence virtually dies out by $L \sim 8$. From non-equilibrium susceptibility data we obtain an estimate for the value of the dynamic correction to scaling exponent which is large, $w = 2.9(6)$, and very similar to a previous estimate of the static correction to scaling exponent $\omega$ \cite{pom}.

From an analysis of non-equilibrium data on large samples together with equilibrium susceptibilites \cite{pom,lwb}, our estimates for $T_g = 1.195(15)$, and the associated static critical exponents, are in excellent agreement with those of  Ogielski \cite{og} who made dynamic and static measurements on even larger samples (up to $L=64$). They are close to those of Palassini and Caracciolo \cite{pala}, who carefully extrapolated to infinite size, explicitly allowing for corrections to finite size scaling. They are consistent with high quality series results \cite{singh}, which come from a method complementary to and independent of the Monte Carlo numerical simulations.  When Binder parameter data are extrapolated to large $L$, the $T_g$ estimate obtained is imprecise because of uncertainties in the corrections to finite size scaling, but within the errors it is consistent with the estimates using the other techniques. There is however a puzzling residual disagreement with the work of Ballesteros et al \cite{Bal} concerning the precise value of the critical temperature $T_g$.

We gratefully acknowledge the use of the computing facilities of IDRIS.



\begin{references}



\bibitem{EA} S.F. Edwards and P.W. Anderson, J. Phys. F {\bf 5}, 965 (1975)
\bibitem{ground} M. Palassini and A.P. Young, Phys. Rev. B {\bf 60}, 9919 (1999),
J. Houdayer and O.C. Martin, Europhys. Lett. {\bf 49}, 794 (2000),
E. Marinari and G. Parisi, Phys. Rev. B {\bf 62}, 11677 (2000)
\bibitem{BY} R.N. Bhatt and A.P. Young, Phys. Rev. B {\bf 37}, 5606 (1988)
\bibitem{og} A.T. Ogielski, Phys. Rev. B {\bf 32}, 7384 (1985)
\bibitem{KY} N. Kawashima and A.P Young, Phys. Rev. B {\bf 53}, R484 (1996)
\bibitem{pala} M. Palassini and Caracciolo, Phys. Rev. Lett. {\bf 82}, 5128 (1999)
\bibitem{pom} P.O. Mari and I.A. Campbell,  Phys Rev E {\bf 59}, 2653 (1999)
\bibitem{Bal} H.G. Ballesteros, A. Cruz, L.A. Fern\'andez, V. Mart\'in-Mayor, J. Pech, J.J. Ruiz-Lorenzo, A. Taranc\'on, P. T\'ellez, C.L. Ullod and C. Ungil, Phys. Rev. B {\bf 62},
14237 (2000)
\bibitem{cruz} A. Cruz, J. Pech, A. Tarnc\'on, P. T\'ellez, C.L. Ullod and C. Ungil, cond-mat/0004080, Comp. Phys. Comm. {\bf 133}, 165 (2000)
\bibitem{singh} R.R.P. Singh and S. Chakravarty, Phys. Rev. B {\bf 36}, 559 (1987)
\bibitem{d4} G. Parisi, F. Ricci-Tersenghi and J.J. Ruiz-Lorenzo, Phys. Rev. B {\bf 57}, 13617 (1998)
\bibitem{lwb} L.W. Bernardi, S. Prakash and I.A. Campbell, Phys. Rev. Lett. {\bf 77}, 2798 (1996)
\bibitem{huse} D.A. Huse, Phys. Rev. B {\bf 40}, 304 (1989)
\bibitem{rieger} H. Rieger, J. Phys. A {\bf 26}, L615 (1993)
\bibitem{zheng} B. Zheng, M. Schulz and S. Trimper, Phys. Rev. E {\bf 59}, R1351 (1999)
\bibitem{parisi} G. Parisi, F. Ricci-Tersenghi and J.J. Ruiz-Lorenzo, Phys. Rev. E {\bf 60}, 5198 (1999)
\bibitem{ma}S.K. Ma, {\it Modern theory of Critical Phenomena} (Benjamin, Reading MA, 1976)
\bibitem{bray} R.E. Blundell, K. Humayun and A.J. Bray, J. Phys. A {\bf 25}, L733 (1992)
\bibitem{klein} L. Klein, J. Adler, A. Aharony, A.B. Harris and Y. Meir, Phys. Rev. B {\bf 43}, 11249 (1991), J. Adler, Annu. Rev. Comput. Phys. {\bf 4}, 241 (1996)
\bibitem{pelissetto} A. Pelissetto and E. Vicari, cond-mat/0012164
\bibitem{perco} H.G. Ballesteros et al, J. Phys. A {\bf 32}, 1 (1999)
\bibitem{marinari} E. Marinari, G. Parisi and F. Ritort, J. Phys A {\bf 27}, 2687 (1994)
\bibitem{jerome} J. Houdayer, Eur. Phys. J. B {\bf 22}, 479 (2001)
\bibitem{endo} E. Marinari and F. Zuliani, J. Phys A {\bf 32}, 4774 (1999)
\bibitem{ian} I.A. Campbell, D. Petit, P.O. Mari and L.W. Bernardi J. Phys. Soc. Jpn {\bf 69} Suppl. A, 186 (2000)
\bibitem{adler} J. Adler et al, Phys. Rev. B {\bf 41}, 9183 (1990)

\end{references}
\end{document}